\documentstyle[pra,aps,multicol,graphicx,subfigure]{revtex}
\begin{document}

\title{Vortex Penetration  In Magneto-Superconducting 
Heterostructures.}

\author{Serkan Erdin\\
\vspace{0.5cm}
School of Physics \& Astronomy, University of 
Minnesota,\\
116 Church St. S.E., Minneapolis, MN 55455}

\maketitle

\begin{abstract}
We report our results on vortex penetration
in two realizations of heterogeneos magneto-superconducting 
systems (HMSS) based on London approach; semi-infinite 
ferromagnetic(FM)-superconducting(SC) bilayers
and a FM dot on a semi-infinite SC
film. In the first case, 
we study 
quantitatively the vortex entry in FM-SC bilayers which manifests 
Bean-Livingston-like vortex barrier, 
controlled by FM film's magnetization 
$m$ and SC film's Ginzburg 
parameter $\kappa$. In the second case, we investigate  the conditions 
for sponteneous vortex creation  and determine the position of vortex for 
various values of magnetization and the dot's position. 
\end{abstract}

\vspace{1cm}

\noindent PACS Number(s): 74.25.Dw, 74.25.Ha, 74.25.Qt, 74.78.-w

\vspace{1cm}

\begin{multicols}{2}

Heterogeneous magneto-superconducting systems (HMSS) are  made of 
ferromagnetic (FM) 
and superconducting
(SC) pieces separated by thin layers of insulating oxides. In contrast
to the case of a homogeneous ferromagnetic superconductor studied during
the last two
decades, the two order parameters, the magnetization and the SC electron
density do not suppress each other \cite{pok1,pl1}.
In HMSS, the strong interaction between FM and SC components stems 
from the
magnetic fields generated by the inhomogeneous magnetization and
the supercurrents
as well as SC vortices. Strong
interaction of the FM and SC systems not only gives rise to a new
class
of novel phenomena and physical effects, but also shows the important
technological promise of devices whose transport properties can be easily
tuned by comparatively weak magnetic fields.
 
Various theoretical  realizations of HMSS have been proposed by 
different groups,  such as
arrays of
magnetic dots on the
top of a SC film \cite{pok1,se1}, ferromagnetic/superconducting bilayers (FSB)
\cite{pok2}, and magnetic nanorods embedded into a superconductor
\cite{pok3}, whereas only sub-micron magnetic dots
covered by thin SC films have been prepared and studied
\cite{e1,e2,e4}. The experimental samples of FM-SC hybrid
systems were prepared by means of
electron beam lithography and lift-off techniques \cite{ee}. Both 
in-plane
and
out-of-plane magnetization was experimentally studied. The dots with
magnetization parallel to the plane were  fabricated from
Co, Ni, Fe, Gd-Co and Sm-Co alloys. For the dots with
magnetization
perpendicular to the plane which requires high anisotropy along hard-axis, 
Co/Pt multilayers were used \cite{vanbael}. 

In 
the most of 
theoretical 
studies, SC subsystem is considered to be 
infinite size for the sake of computational simplicity.
To this 
date,
there has not been an analytical analysis of boundary and edge effects in
FM-SC heterostructures, though vortex entry
conditions  in type II superconductors   are previously studied
\cite{degennes,kramer,brandt}.
However,
from both experimental and theoretical point of view, finite or 
semi-infinite systems
are
more interesting and realistic, and their study offers better 
understanding of vortex matter in HMSS. 
Author also believes that analytical and quantitative study of 
aforementioned 
systems
will shed light on solving other open problems pertaining to HMSS. For example, 
we earlier predicted that in a finite
temperature interval
below the SC
transition the
FSB is unstable with respect to SC vortex formation in FM-SC bilayers
(FSB) 
\cite{domains}.  The 
slow decay
($\propto 1/r$) of the long-range interactions between Pearl
vortices makes the structure that consists of alternating domains with
opposite magnetization and vorticity energetically favorable. 
It is possible that the long domain nucleation time can 
interfere with
the observation of described textures. We also expect that domain 
nucleation 
starts near the edge, which makes qualitative study of edges in 
aforementioned systems necessary.
Quantitative study of this dynamic process is still in progress.
For this purpose, and having been motivated by current interest in HMSS, 
in this work, we attempt to study vortex entry conditions in HMSS. 
To our purpose, we work 
with a method based on London-Maxwell equations, which is fully explained 
elsewhere \cite{th4}. London
approach
works
well for large Ginzburg ($\kappa = \lambda_{eff}/\xi  >> 1$ ) parameter, 
where $\lambda_{eff} = \lambda_L^2/d$ is the effective 
penetration depth \cite{abrikosov}, and $\xi$ is coherence length.
Indeed,
for thin SC films, Ginzburg  parameter is on order of 50-100. 
Previously, our method was introduced for   vortex structures in 
infinite films. Here, we extend it to semi-infinite systems. To
this end, we benefit from Kogan's work on a Pearl vortex near the edge of 
SC thin film in which SC piece's size is considered to be semi-infinite 
\cite{kogan}. 
Likewise, we
consider FM subsystems on 
semi-infinite SC and FM subsystems in which, we assume 
that magnetization points perpendicular to the FM film's plane.

In this work, we first consider semi-infinite SC and FM films and study 
vortex entry
barrier. Our calculations show that
there exists Bean-Livingston-like surface barrier \cite{bean} for the 
vortices created by FM
film.  Next, we consider a circular magnetic dot near the
film's edge and investigate the conditions for vortices to appear and
their configurations. It turns out that, in contrast to the infinite
systems, vortices are not trapped right at the dots center, but they are
shifted  slightly from the center to the SC film's edge or opposite
direction, depending on the dot's magnetization, position and size.
Physics behind this effect is simple. In the semi-infinite systems, 
vortex interacts with both its image vortex and the magnetic dot. 
 The competition between these two 
attractions determines vortex's position.
The outline of this articles is as follows: in the first section, we 
introduce the method to study edge effects in FM-SC systems. In the next 
section, we apply our method based on Maxwell-London equations to two 
different cases; semi-infinite FM-SC bilayers and FM dot on a 
semi-infinite SC film. We conclude with results and discussions.

\section{Method}

\begin{figure}[h]
\begin{center}
\fbox{\includegraphics[angle=0,width=3.0in]{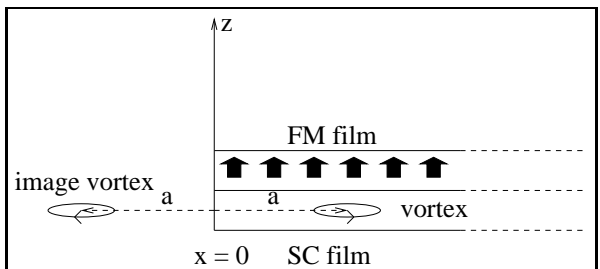}}
\caption{\label{semi} Semi-infinite FM-SC bilayer.}
\end{center}
\end{figure}

\noindent Finite and semi-infinite 
systems are not as easy and straightforward as infinite systems, and they 
usually require more careful treatment due to the boundary of the systems. 
Earlier, Kogan 
developed a clever technique based on London approach to study a vortex 
near 
the 2d film's edge \cite{kogan}. While developing our method, we stick to 
his 
technique and geometry in which a very thin SC film is located at x-y 
half plane while its edge is at x=0 (see Fig.\ref{semi}), and generalize 
his method for more 
than one vortex. We also 
assume that no vortex is closer to the SC film's edge than coherence 
length $\xi$, because London theory does fail in the vicinity of $\xi$.

\noindent We start with London equation for the vortices with vorticity 
$n_j$ 
located at ${\bf 
r}_j$, 

\begin{equation}
{\bf h} + \frac{4 \pi \lambda_L^2}{c} {\bf \nabla} \times {\bf j}_s = 
\phi_0 
\hat z \sum_j n_j \delta ({\bf r} - {\bf r}_j), 
\label{eq1}
\end{equation}

\noindent where ${\bf j}_s$ is supercurrent density in the SC subsystem. 
In the presence of FM subsystem which is also considered to be very thin 
and located at x-y half plane as SC subpiece, Eq.(\ref{eq1})  
turns to

\begin{equation}
{\bf h} + \frac{4 \pi \lambda_L^2}{c} {\bf \nabla} \times {\bf j} = 
\frac{4 
\pi \lambda_L^2}{c} {\bf \nabla} \times {\bf j}_m + \phi_0 \hat z 
\sum_j n_j \delta ( 
{\bf r} - {\bf r}_j ),
\label{eq2}
\end{equation}

\noindent where ${\bf j} = {\bf j}_s + {\bf j}_m$ and ${\bf j}_m = 
c {\bf \nabla} \times
{\bf m}$. Averaging Eq.(\ref{eq2}) over the thickness of SC film, one 
finds  

\begin{equation}
h_z +  \frac{4 \pi \lambda}{c} ({\bf \nabla} \times {\bf g})_z =
\biggl ( \frac{4
\pi \lambda}{c} {\bf \nabla} \times {\bf g}_m \biggr )_z + \phi_0 \sum_j 
n_j \delta ( 
{\bf 
r} - {\bf r}_j),
\label{eq3}
\end{equation}
\noindent where ${\bf g}$ is the 2-d current density, which can be 
calculated by solving 
Eq.(\ref{eq3}) together with the Biot-Savart integral equation and the 
continuity equation ${\bf \nabla} \cdot {\bf g} = 0$. In terms of the 
surface current, the Biot-Savart equation  is given by

\begin{equation}
h_z = \frac{1}{c}\int d^2 {\bf 
r}^\prime [ {\bf g} ( {\bf r}^\prime ) \times {\bf R}/R^3 ], 
\label{eq4}
\end{equation}
\noindent where ${\bf R} = {\bf r} - {\bf r}^\prime$. Defining the surface 
current density in terms of a scalar function $G({\bf r})$ as ${\bf g} = 
{\bf \nabla} \times G({\bf r}) \hat z$, using ${\bf R}/R^3 = {\bf 
\nabla}^\prime (1/R)$ and integrating Eq.(\ref{eq4}) by parts,  one can 
obtain

\begin{equation}
h_z = \frac{1}{c} \int_{x^\prime > 0} d^2 {\bf r}^\prime \frac{\nabla^2 G 
({\bf
r}^\prime)}{R} + \int_{-\infty}^\infty d y^\prime \left( \frac{
\partial_{x^\prime} G ( {\bf
r}^\prime)}{R} \right)_{x^\prime = 0}, 
\label{eq5}
\end{equation}

\noindent where the first term gives the contribution from the entire 
surface 
current distribution whereas the second term is the contribution from the 
film's edge. The direct substitution of Eq.(\ref{eq5}) into Eq.(\ref{eq2}) 
gives 

\end{multicols}

\begin{equation}
 \int_{x^\prime > 0} d^2 {\bf r}^\prime \frac{\nabla^2 G ({\bf 
r}^\prime)}{R} + \int_{-\infty}^\infty d y^\prime \left( \frac{ 
\partial_{x^\prime} G ( {\bf
r}^\prime)}{R} \right)_{x^\prime = 0} + 4 \pi \lambda \nabla^2 G ({\bf r}) 
= - 
c \phi_0 \sum_j n_j
\delta ( {\bf r} - {\bf r}_j  )  
- 4 
\pi \lambda  ({\bf \nabla} \times {\bf g}_m)_z. 
\label{eq6}
\end{equation}

\begin{multicols}{2}
\noindent Solving the above equation in half plane is difficult. However, 
this difficulty can be removed by  solving (\ref{eq6}) in the Fourier 
space and using  the boundary conditions. That is, at the films's edge ( 
$x = 0$ ), the normal component of current density is zero, namely $g_x ( 
0 , y ) = 0$, 
whereas, at infinity
the current distribution vanishes.  This implies that the scalar function
is constant at the films boundaries. For simplicity, it can be set to 
zero. 
To have G vanish at the edge, we set $G(-x,y)= -G(x,y)$. 
  The Fourier transform of Eq.(\ref{eq6}) reads

\end{multicols}

\begin{eqnarray}
\label{eq7}
\int_0^\infty dx^\prime e^{- i k_x x^\prime} ( \partial^2_{x^\prime} - 
k_y^2 ) G ( x^\prime, k_y ) &+& \int_{-\infty}^\infty d y^\prime [ 
\partial_{x^\prime} G ({\bf r}^\prime ) ]_{x^\prime=0} e^{- i k_y 
y^\prime} - 2 
\lambda k^3 G({\bf k}) \\ \nonumber 
&=& i \frac{c \phi_0}{\pi} k \sum_j n_j e^{-i k_y y_j} \sin ( k_x 
x_j ) - 2 \lambda k ( i {\bf k} \times {\bf g}_{m, k_x} )_z, 
\end{eqnarray}

\begin{multicols}{2}

\noindent where ${\bf k} = (k_x,k_y)$. Replacing 
$x^\prime$ by $-x^\prime$ and writing the Eq.(\ref{eq7}) for $-k_x$, 
one obtains

\end{multicols}

\begin{eqnarray}
\label{eq8}
\int_0^{-\infty} dx ^\prime e^{- i k_x x^\prime} ( \partial^2_{x^\prime} -   
k_y^2 ) G ( x^\prime, k_y ) &+& \int_{-\infty}^\infty d y^\prime [
\partial_{x^\prime} G ({\bf r}^\prime ) ]_{x^\prime=0} e^{- i k_y 
y^\prime} + 2
\lambda k^3 G({\bf k}) \\ \nonumber
&=& - i \frac{c \phi_0}{\pi} k \sum_j n_j e^{-i k_y y_j}  \sin ( k_x x_j 
) - 2 
\lambda k ( i \tilde 
{\bf 
k} 
\times {\bf g}_{m, - k_x} )_z, 
\end{eqnarray}
 
\begin{multicols}{2}

\noindent where $\tilde {\bf k} = (-k_x,k_y)$.  Subtracting (\ref{eq8}) 
from (\ref{eq7}), the vortex and 
magnetic parts of the scalar function are found as

\begin{eqnarray}
G_v ({\bf k}) &=& \frac{2 c \phi_0}{i \pi}\sum_j n_j \frac{e^{-i k_y 
y_j} \sin 
( k_x x_j )}{k ( 1 + 4 
\lambda k )},\label{eq9} \\
G_m ({\bf k}) &=&  2 \lambda  i \frac{[ {\bf k} \times {\bf g}_{m, k_x} 
- 
\tilde {\bf
k}
\times {\bf g}_{m, - k_x} ]_z}{k ( 1 + 4 \lambda k )}.
\label{eq10}
\end{eqnarray}

\noindent Taking the inverse Fourier transform of Eq.(\ref{eq9}), the 
vortex contribution in 
real space is found as

\begin{equation}
G_v({\bf r}) = \frac{c \phi_0}{2 \pi^2} \sum_j  n_j \int_0^\infty 
\frac{(J_0 ( 
k 
|{\bf 
r}- {\bf r}_j|) - J_0 ( k |{\bf
r}+ \tilde {\bf r}_j|))}{1 + 4 k \lambda} d k, \label{eq11} 
\end{equation}

\noindent Note that the first term in (\ref{eq11}) represents the $j^{th}$
vortex located at ${\bf r}_j = (x_j,y_j)$, whereas the 
second term is the contribution of  
$j^{th}$ image vortex, or 
antivortex, at $\tilde  {\bf r}_j  = (-x_j, y_j)$. Next, we 
 calculate the 2-d current density. 
Keeping in mind that it is discontinuous at the film's edge, 
the Fourier components of the current density read

\begin{eqnarray}
{\bf g} ( x > 0, y ) &=& \int \frac{ d^2 {\bf k}}{ ( 2 \pi )^2} ( i {\bf
k}
\times \hat z ) G ({\bf k }) e^{i {\bf k}\cdot {\bf r}}, \nonumber \\
{\bf g} ( x < 0, y ) &=& 0.
\label{eq12}
\end{eqnarray}


\noindent Using Eq.(\ref{eq12}), one can compute vector potential and the 
magnetic field 
through 

\begin{equation}
{\bf A} = \frac{4 \pi}{c} \frac{\bf g}{Q^2}, \hspace{1cm} {\bf h} = 
\frac{4 
\pi}{i c} \frac{{\bf g} \times {\bf Q}}{Q^2}, 
\label{eq13}
\end{equation}
\noindent where ${\bf Q} = {\bf k} + k_z \hat z$. Taking the inverse 
Fourier of Eq.(\ref{eq13}), vector potential is found as

\begin{equation}
 A_\phi ( {\bf r}) = \frac{\phi_0}{\pi}\sum_j n_j  \int_0^\infty 
\frac{(J_1 ( 
k 
|{\bf r} -
{\bf r}_j|) - J_1 ( k |{\bf r} + \tilde {\bf r}_j|)) e^{-k |z|}}{1 + 4 
\lambda k} d k.  
\end{equation}

\noindent At the SC film's surface (z = 0), vector potential for one 
vortex with vorticity n,  located at ${\bf r} = {\bf a}$ reads

\end{multicols}

\begin{equation}
A_\phi ( {\bf r})
= \frac{n \phi_0}{4 \lambda \pi}  \left( \frac{4 \lambda}{|{\bf 
r}-{\bf 
a}|} 
- 
\frac{4 \lambda}{
|{\bf r} + \tilde {\bf a}|} 
+ \frac{\pi}{2} \left[ Y_1 \left(\frac{|{\bf r}-{\bf a}|}{4 
\lambda} \right) + H_{-1} \left(\frac{|{\bf r}-{\bf a}|}{4
\lambda} \right) - Y_1 \left(\frac{|{\bf r}+\tilde {\bf a}|}{4
\lambda} \right) + H_{-1} \left(\frac{|{\bf r}+\tilde {\bf a}|}{4
\lambda} \right) \right] \right), 
\end{equation}

\begin{multicols}{2}

\noindent where  $H$ and $Y$ are the
Struve and the second kind Bessel
fuctions. At short distances ($ r << 
\lambda$), $A_\phi$ behaves as
$(n \phi_0/16 \pi \lambda^2)[ (-1/4+C/2 - \ln2/2 ) ( |{\bf r}-{\bf 
a}| 
- 
|{\bf r}+{\bf a}|) +  |{\bf r}-{\bf a}| \ln( |{\bf r}-{\bf a}|/4 \lambda) 
-|{\bf r}+{\bf a}| \ln( |{\bf r}+{\bf a}|/4 \lambda)$ whereas, at large 
distances, it decays slowly in space, namely
$ (\phi_0/\pi)(1/|{\bf r}-{\bf a}| - 1/|{\bf r}+{\bf a}|)$. $C=0.577...$ 
is Euler constant.
Magnetic field due to vortex in z direction reads

\begin{equation}
h_z ({\bf r}) = \frac{n \phi_0}{\pi} \int_0^\infty \frac{(J_0 ( k |{\bf 
r} - 
{\bf a}|) - J_0 ( k |{\bf r} + {\bf a}|)) k e^{-k |z|}}{1 + 4 \lambda k} 
d k 
\end{equation}

\noindent At z = 0, magnetic field reads

\end{multicols}

\begin{equation}
h_z ({\bf r})= \frac{n \phi_0}{4 \lambda \pi}\left(\frac{1}{|{\bf r}-{\bf 
a}|} 
-
\frac{1}{|{\bf r} + {\bf a}|} 
- \frac{\pi}{8 \lambda} \left[ H_{0} \left(\frac{|{\bf 
r}-{\bf a}|}{4
\lambda} \right) - Y_0 \left(\frac{|{\bf r}-{\bf a}|}{4
\lambda} \right) - H_{0} \left(\frac{|{\bf r}+{\bf a}|}{4
\lambda} \right) - Y_0 \left(\frac{|{\bf r}+{\bf a}|}{4
\lambda} \right) \right] \right).
\label{eq14}
\end{equation}

\begin{multicols}{2}

\noindent The asymptotics of the magnetic field at small and large 
distances are

\begin{eqnarray}
h_z &\sim&  \frac{n \phi_0}{4 \pi \lambda} \left( \frac{1}{|{\bf r}-{\bf 
a}|} 
-
\frac{1}{|{\bf r}+{\bf a}|} + \frac{1}{4 \lambda} \ln \frac{|{\bf r}-{\bf 
a}|}{|{\bf r}+{\bf a}|} \right) 
\hspace{0.2cm} r << \lambda \\
h_z &\sim&   \frac{4 n \lambda \phi_0}{\pi} \left( \frac{1}{|{\bf 
r}-{\bf a}|^3} -
\frac{1}{|{\bf r}+{\bf a}|^3}\right)\hspace{1cm} r >> \lambda.
\end{eqnarray}

\noindent The total  energy of FM-SC system   reads
\begin{equation}
E = E_v + E_{vm} + E_m,
\end{equation}
where $E_v$ is the vortex energy, $E_{vm}$ is the interaction of 
vortex and magnetic subsystem and finally $E_m$ is selfenergy of 
magnetic subsystem, which will be ignored at further calculations, 
since it is inappropriate for our problem. 
Vortex energy is 
calculated by Kogan \cite{kogan} as

\begin{equation} 
E_v = \sum^{\prime}_{i,j} \frac{n_i \phi_0}{2 c}  G_v ( {\bf r} 
\rightarrow 
{\bf 
r}_j),\label{ev}
\end{equation}

\noindent  where $\sum^{\prime}$  denotes the restricted sum in which 
only $i=j$ and $i>j$ are taken into account. Eq.(\ref{ev})  leads to

\end{multicols}

\begin{equation}
E_v = \sum_i \frac{n_i^2  \phi_0^2}{16 \pi^2 \lambda} \left[ \ln \frac{8 
\lambda}{e^C \xi} - \frac{\pi}{2}  
\Phi_0 \left(\frac{ x_i}{2 \lambda} \right) \right] +
\sum_{i>j}  \pi \varepsilon_0 n_i n_j  \left[\Phi_0 \left(\frac{{\bf 
r}_i -{\bf r}_j}{4 \lambda} \right) - \Phi_0 \left(\frac{{\bf
r}_i +{\bf r}_j}{4 \lambda} \right) \right],\label{voren}
\end{equation}

\begin{multicols}{2}

\noindent where $\Phi_0 ( x ) = Y_0 ( x ) - H_0 ( x )$. 
Vortex-magnetization interaction 
energy is calculated  as in \cite{th4}: 

\begin{equation}
E_{vm} = -\frac{\phi_0}{16 \pi \lambda^2} \int {\bf \nabla} 
\varphi \cdot {\bf a}^m d^2 x - \frac{1}{2} \int {\bf m} \cdot 
{\bf b}^v d^2 x,\label{vormagen}
\end{equation}

\noindent where integration is performed over the half space. Note that
we take $\lambda/\xi = 50$ in our numerical calculations.

\subsection{Semi-Infinite FM-SC Bilayers}

In this part, we study a semi-infinite FM film on top 
of a semi-infinite SC film. 
Both films are taken to be very 
thin, lie on x-y half plane whereas their edges are located at x = 0 (see 
Fig.\ref{semi}). We assume that FM film has uniform 
magnetization along z direction and has high anisotropy, so that its 
magnetization does not change direction due to magnetic field of 
vortex. The magnetization of FM film reads

\begin{equation}
{\bf m} = m \theta ( x)  \delta(z) \hat z.\label{mag}
\end{equation}

\noindent Magnetic current in real space and Fourier space is given as

\begin{equation}
{\bf g}_m = - m c \delta ( x ) \hat y, \hspace{1cm} {\bf 
g}_{m,k} = - 2 
\pi m 
c \delta (k_y) \hat k_y. \label{magcur} 
\end{equation}
 
\noindent Substituting Eq.(\ref{mag})    into Eq.(\ref{eq10}) , one can 
find the 
scalar potential as 

\begin{equation}
G_m ({\bf k}) = -8 \pi \lambda m c i \frac{k_x \delta(k_y) }{k 
( 1 + 4 
\lambda k )}. \label{scal}
\end{equation}

\noindent Taking the inverse Fourier transform of Eq.(\ref{scal}), we 
find 

\begin{equation}
G_m ( x ) = \frac{ m c}{\pi} f\left(\frac{x}{4 \lambda}\right),
\end{equation}
\noindent where $f(x) = \int_{0}^{\infty} d k_x  \sin(k_x 
x)/(1 + 
k_x))$. The asymptotics of $f(x)$ are
\begin{eqnarray}
f(x) &\approx& \frac{\pi}{2} + x ( \ln (x) + C - 1 ), x << 1, \\ \nonumber 
f(x) &\approx& \frac{1}{x}, x >> 1.
\end{eqnarray}


\noindent Using Eq.(\ref{eq12}) and Eq.(\ref{eq13}), z component of the 
screened magnetic field at z = 0 due to FM film, reads

\begin{equation}
h_z ( {\bf r} ) = \frac{m}{4 \lambda} \int_{0}^{\infty} \frac{k_x \sin ( 
k_x \frac{x}{4 \lambda})}{1 + |k_x|} d k_x. 
\end{equation}

\noindent The magnetic field decays as $1/x$ for $x << \lambda$ and 
$1/x^2$ for $x >> \lambda$. In order to study vortex configuration, we 
need to calculate 
total effective energy of the system. To this end, we consider a simple 
case, namely a vortex with a single  flux located at ${\bf r} =
{\bf a}$. For this case, vortex energy for a single vortex reads (see 
Eq.(\ref{voren}))

\begin{equation}
E_v = \frac{\phi_0^2}{16 \pi^2 \lambda} \left[ \ln \frac{8
\lambda}{e^\lambda \xi} - \frac{\pi}{2}
\Phi_0 \left(\frac{a}{2 \lambda} \right) \right], \label{eq30} 
\end{equation}
 
\noindent wheras the vortex-magnetization interaction energy can be 
calculated by means of Eq.(\ref{vormagen}). 
\begin{equation}
E_{vm} = - m \phi_0 \left[ 1 - \frac{2}{\pi} f\left(\frac{a}{4 
\lambda}\right ) \right]. 
\label{evm}
\end{equation}

\begin{figure}[h]
\begin{center}
\fbox{\includegraphics[angle=270,width=3.0in]{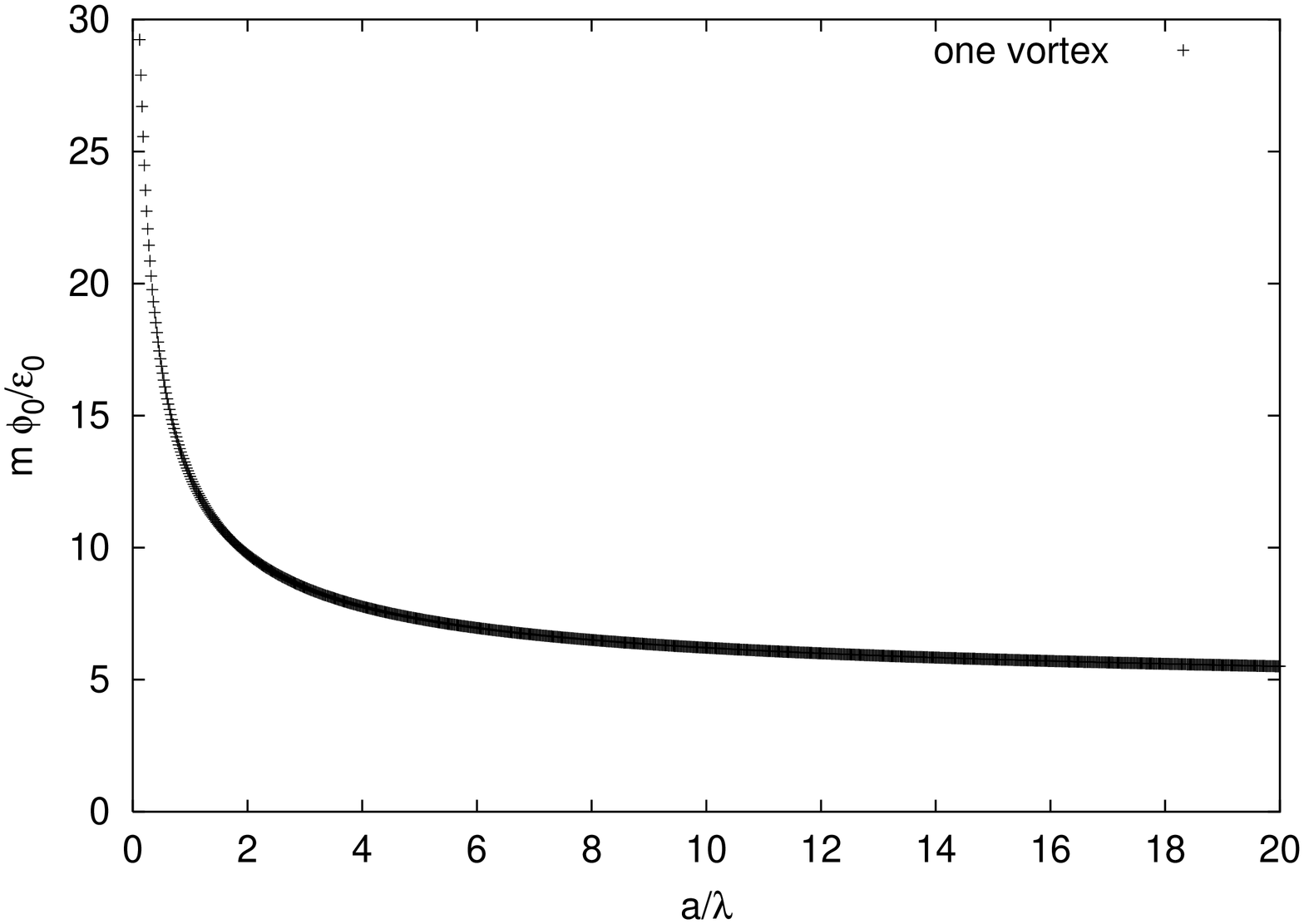}}
\caption{\label{1vor} Phase diagram of a single vortex created by 
semi-infinite FM film. In the region below curve, vortex does not 
appear, whereas it becomes energetically favorable in area above the 
curve.} 
\end{center} 
\end{figure}

\begin{figure}[h]
\begin{center}
\fbox{\includegraphics[angle=270,width=3.0in]{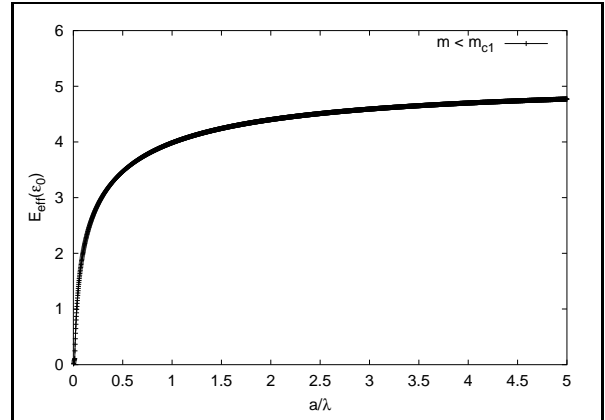}}
\caption{\label{barrier1} The effective energy versus the vortex's 
position. When $m < m_{c1}$, vortex does not appear.}
\end{center}
\end{figure}

\begin{figure}[h]
\begin{center}
\fbox{\includegraphics[angle=270,width=3.0in]{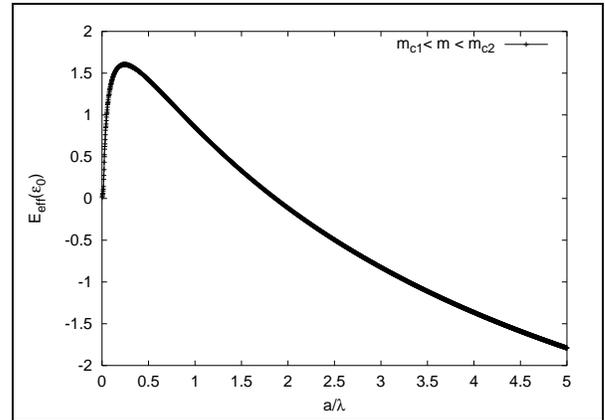}}
\caption{\label{barrier2} The effective energy versus the vortex's
position. When $m_{c1} < m < m_{c2}$, the surface barrier shrinks toward 
the edge 
of SC film, and vortex is created little further from the edge.}
\end{center}
\end{figure}

\noindent The sum of  Eq.(\ref{voren}) and Eq.(\ref{evm}) gives the 
effective total energy of the system. Vortex becomes energetically 
favorable when effective total energy becomes less than zero. Equating the 
effective energy to zero, one can obtain the curve for spontenous creation 
of vortex. This curve, 

\begin{equation}
\frac{m \phi_0}{\varepsilon_0} = \frac{\ln (\frac{8
\lambda}{e^\lambda \xi}) - \frac{\pi}{2} ( H_0 ( \frac{a}{2
\lambda}) - Y_0 (\frac{a}{2
\lambda}))}{1 - \frac{2}{\pi} f ( \frac{a}{4 \lambda})}
\end{equation}

\noindent seperates the regions where the vortex appears spontaneously and 
does appear as seen in 
Fig.\ref{1vor}. For  large values of  $m \phi_0/\varepsilon_0$ ratio,
the vortex comes out  near the edge. On the other 
hand, it 
prefers going further away from the surface for small ratio of $m 
\phi_0/\varepsilon_0$. We can estimate the minimum  value of magnetization 
of FM film through effective energy for infinite films \cite{th4}, which 
is $E_{eff} = \varepsilon_0 \ln ( \lambda /\xi) - m \phi_0$. 
Equating  this 
equation to zero and solving it for $m$, we find $m_{c1} = 
\phi_0/(16 \pi^2 \lambda\ln ( \lambda/\xi))$. When magnetization exceeds 
this value, 
the vortex appears very far away from the edge. In order to get vortex 
appear close to the edge, $m$ must be significantly larger than $m_{c1}$.  
Another 
interesting thing is that the system manifests Bean-like surface barrier 
for the vortex. The surface barrier is controlled by $m 
\phi_0/\varepsilon_0$ and Ginzburg parameter $\kappa$. We analyze three 
regimes for this ratio for fixed $\kappa =\lambda/\xi$. 
When $m < 
m_{c1}$, vortex does not appear (see Fig.\ref{barrier1}). In the second 
regime, $m_{c1} < m < 
m_{c2}$, vortex prefers going further away from the surface, whereas, 
when $m > m_{c2}$, the barrier disappears (see Fig.\ref{barrier2}). 
$m_{c2}$ is 
the second critical 
magnetization, at which barrier disappears, and can be calculated through 
the condition that the slope  
$|\partial E_{tot}/\partial x|_{x = \xi}$ is zero, which 
gives $m 
\phi_0/\varepsilon_0 \approx 2 \pi \kappa/\ln(4 \kappa)$. When ratio $m
\phi_0/\varepsilon_0$ is greater than this , the barrier disappears. 
Physically, 
two contributions play an important role for the vortex barrier. Namely, 
the vortex is attracted 
to SC film's edge through its attraction towards image  vortex  whereas it is 
repelled by FM film's edge. Competition between these two factors controls the barrier.

\begin{figure}[h]
\begin{center}
\fbox{\includegraphics[angle=270,width=3.0in]{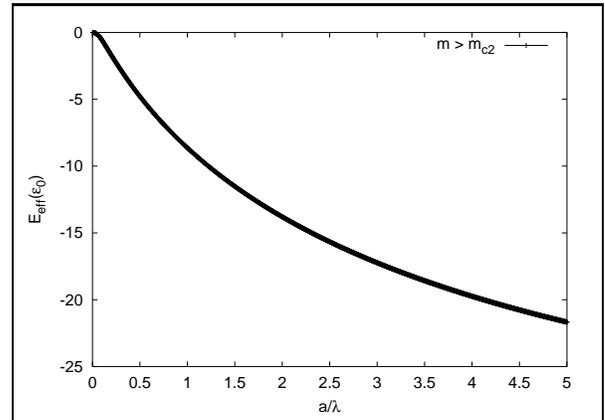}}
\caption{\label{barrier3} The effective energy versus the vortex's
position. When $m > m_{c2}$ barrier disappears, and vortex can be seen 
anywhere in SC film.}
\end{center}
\end{figure}
 





\subsection{FM Dot on a Semi-Infinite SC Film}
In this case, we study a circular FM disc on top of semi-infinite
SC film. Earlier, we studied the conditions for the vortex states to 
appear on a similar system, in which SC film, however was infinite. Due 
to the circular symmetry of FM dot, vortex was appearing at the dot's 
center. In this section, we study the vortex states in more realastic case 
and investigate the role of edge effects on the spontaneous formation of 
vortex due to FM dot. To this end, we start with the magnetization of 
a circular FM disc located 
at ${\bf r}_d = ( x_d , 0)$, 
\begin{equation}
{\bf m} = m \theta (R - |{\bf r} - {\bf r}_d|) \delta (z) \hat z,
\label{mdot}
\end{equation}
\noindent where $R$ is the radius of circular dot. The Fourier transform 
of Eq.(\ref{mdot})   reads
\begin{equation}
{\bf m}_{\bf k} = 2 \pi m R \frac{J_1 ( k R ) e^{-i k_x 
x_d}}{k}.\label{fmdot}
\end{equation} 
\noindent Using Eq. (\ref{eq10},\ref{eq12},\ref{eq13}) together 
with 
Eq.(\ref{fmdot}), one can calculate the screened magnetic field due to FM 
dot as
\end{multicols}

\begin{equation}
h_z (r,z) = 4 \pi m \lambda R \int_0^{\infty} \frac{J_1 ( k R ) [ J_0 ( k 
|{\bf r} - {\bf r}_d|) -J_0 ( k |{\bf r} + {\bf r}_d|)]k^2 e^{-k |z|}}{1 
+ 4 \lambda k} d k.
\end{equation}

\begin{multicols}{2}

\noindent Outside the dot, magnetic field decays rapidly in space, 
namely,
for $r << \lambda$, $\sim 1/r^3$, whereas for $r << \lambda$, $\sim
1/r^5$.

\noindent From Eq.(\ref{vormagen}), the vortex-magnetic disc interaction 
energy reads

\end{multicols}

\begin{equation}
E_{vm} = - m \phi_0 R \int_0^\infty J_1 ( k R ) \frac{[J_0 ( k 
|{\bf r}_d  - {\bf a}|)  - J_0 ( k
|{\bf r}_d +  {\bf a}|)]}{1 + 4 \lambda k} d k. 
\end{equation}

\begin{figure}
\centering
\subfigure[$r_d/\lambda=0.5$] 
{
    \label{fig:sub:a}
    \includegraphics[angle=270,width=7cm]{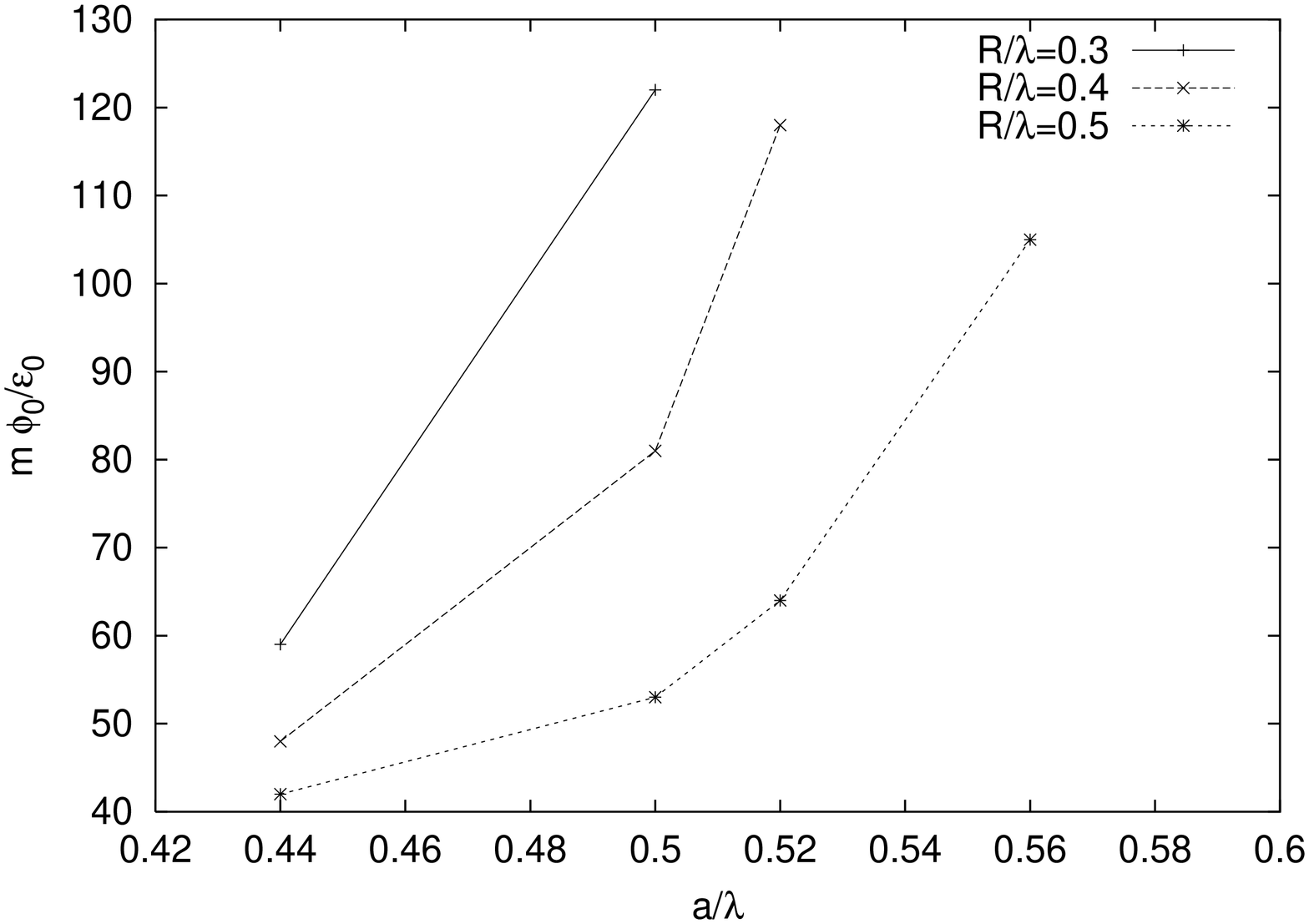}
}
\hspace{1cm}
\subfigure[$r_d/\lambda=0.8$] 
{
    \label{fig:sub:b}
    \includegraphics[angle=270,width=7cm]{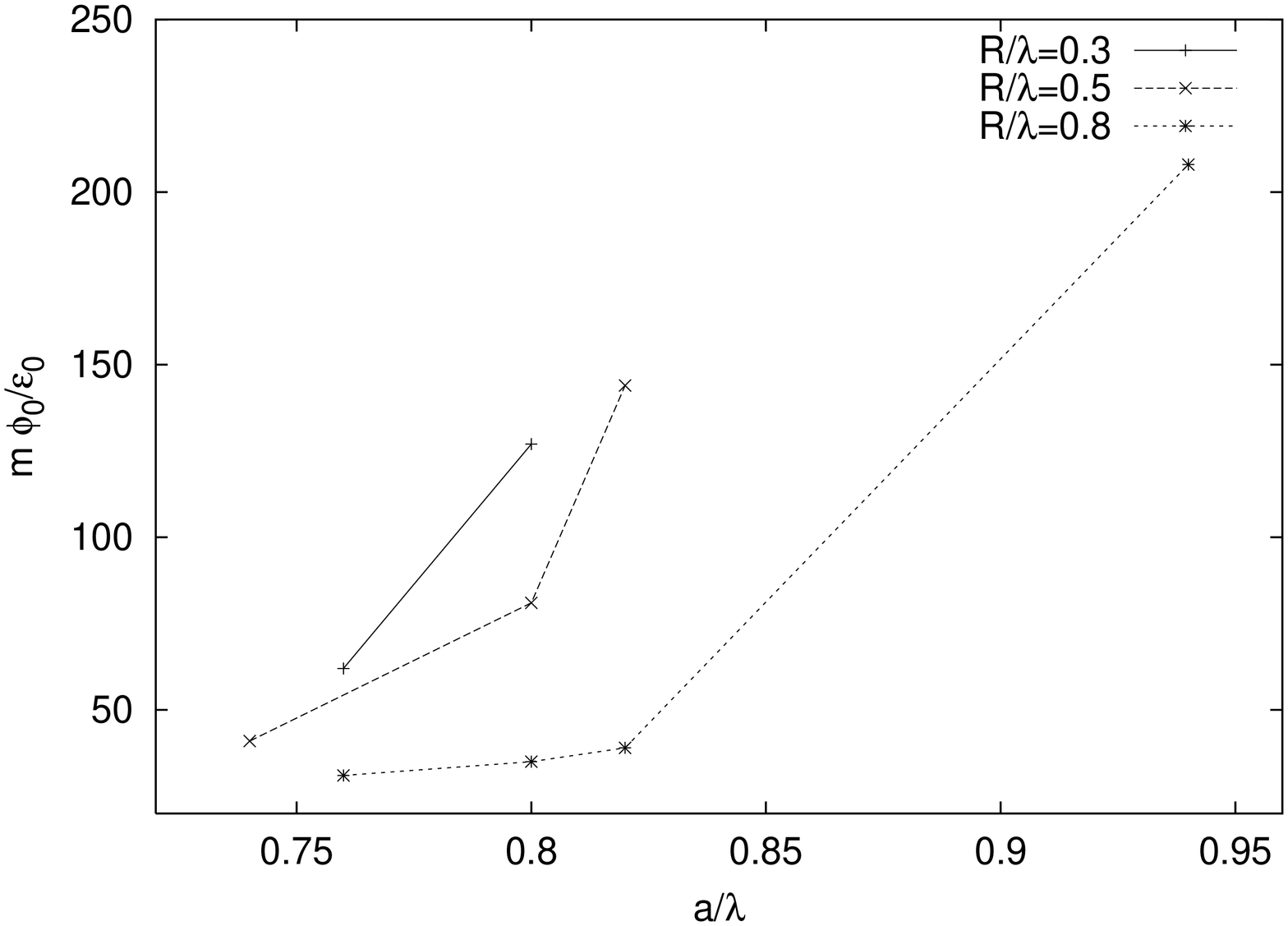}
}
\vspace{1cm}
\subfigure[$r_d/\lambda=1$] 
{
    \label{fig:sub:c}
    \includegraphics[angle=270,width=7cm]{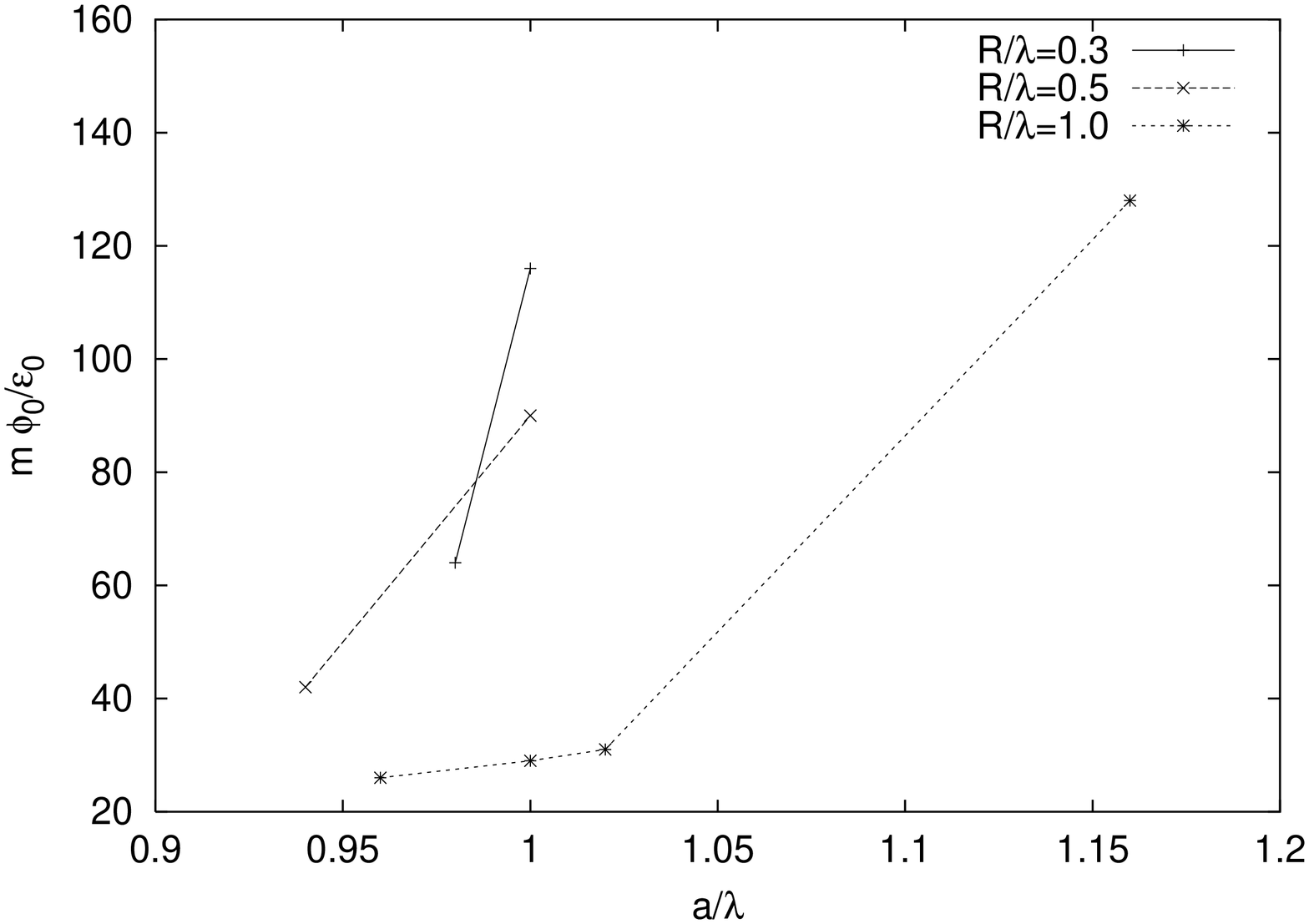}
}
\hspace{1cm}
\subfigure[$r_d/\lambda=5$] 
{
    \label{fig:sub:d}
    \includegraphics[angle=270,width=7cm]{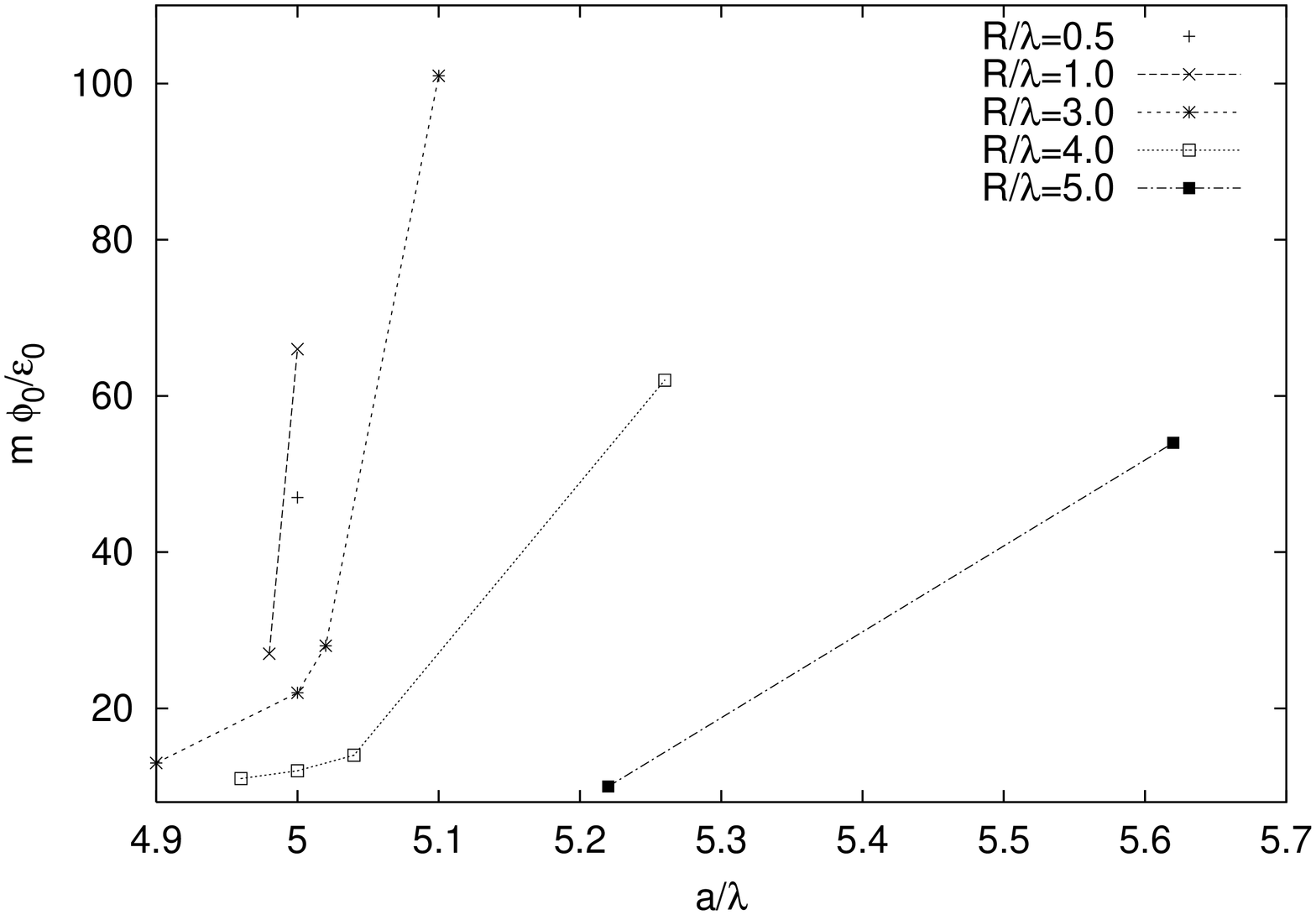}
}
\caption{\label{magndot} Position of vortex $a/\lambda$ versus $m
\phi_0/\varepsilon_0$
for various locations $r_d/\lambda$ and sizes $R/\lambda$ of magnetic
dot}
\end{figure}

\begin{multicols}{2}
\noindent After we formulate the total effective energy as $E= E_v + 
E_{vm}$, where $E_{v}$ is given in Eq.(\ref{eq30}), we study the 
conditions for a vortex to appear spontaneously. The criteria for 
spontaneous vortex formation is that effective energy becomes negative. 
However, vortex in semi-infinite systems also interacts with its image. 
Therefore, it is necessary to minimize total effective energy with respect 
to the vortex position. To this end, we
first fix the dot's location and value of $m\phi_0/\varepsilon_0$ and vary
the vortex's position afterwards, to find the minimum total effective
energy. In our calculations, we investigate where vortex first comes out, 
and how it is shifted with further increase of $m\phi_0/\varepsilon_0$. 
For this purpose, we determine vortex's position for 
different values of the $r_d/\lambda$, $R/\lambda$ and
$m\phi_0/\varepsilon_0$, by optimizing the total effective energy. 
Our results are shown in Table.\ref{vvor} and  
Fig.\ref{magndot}.

According to our calculations, vortex appears first close to the edge
except $r_d/R \geq 10$
case, in which it sits at the dot's center. On the other hand, when
$m\phi_0/\varepsilon_0$ is increased further, vortex is first shifted
towards
the dot's center. With further increase of $m\phi_0/\varepsilon_0$, it
drifts away from the dot's center. However, this is not always
general picture. In the case of  $r_d/R > 5/3$, vortex is located at the
dot's center even for large values of $m\phi_0/\varepsilon_0$.  However,
for larger dot's sizes ($r_d/R \sim 1$ and $r_d/\lambda \geq 2$)
vortex
first appears
away from the dot's center. This situation differs
from the vortex in infinite system. In the latter, only force acting on
vortex stems from the vortex-magnetization interaction, and due to the
dot's circular symmetry, it comes out at the dot's center. However, in
this case, there is another force coming from vortex-image vortex, which
decays slow $\sim 1/r$ for large distances $r > \lambda$ and pulls vortex
towards the SC film's edge, whereas the force exerted by the magnetic dot
pushes vortex towards the dot's center.  As a result,
vortex's position is determined by the balance  between these two forces.

\begin{table}[h]
\caption{The position of vortices for different
values of the $r_d/\lambda$, $R/\lambda$ and $m\phi_0/\varepsilon_0$. The
two columns
on the left are input. \label{vvor}}
\begin{center}
\begin{tabular}{c c c c }
\hline \hline
{$r_d/\lambda$}&{$R/\lambda$}&{$a/\lambda$}&{$m\phi_0/\varepsilon_0$} \\
\hline
2.0 & 2.0 & 2.04 & 17 \\
2.0 & 2.0 & 2.32 & 187 \\
3.0 & 3.0 & 3.12 & 31 \\
3.0 & 3.0 & 3.44 & 132 \\
4.0 & 4.0 & 4.20 & 11 \\
4.0 & 4.0 & 4.56 & 185 \\
\hline
\end{tabular}
\end{center}
\end{table}

\end{multicols}

\begin{multicols}{2}


\section{Conclusions}
In this article, we studied vortex entry conditiions in HMSS. 
First, we generalized Kogan's method for 
quantitative study of semi-infinite HMSS. For applications, we
first considered semi-infinite FM film on top of semi-infinite SC film. 
The quantative analysis of this system showed that the vortex undergoes Bean-like 
barrier which is controlled by two intrinsic properties of system; 
FM film's magnetization $m$ and Ginzburg parameter $\kappa$. Note that our 
result is valid only for a single flux. Our study for the case of several 
fluxes will be published elsewhere.   Secondly, we 
studied 
a single circular  FM dot on a semi-infinite SC film. We analyzed the conditions for 
sponteneous vortex creation and vortex location for various positions of the dot. 
It turns out that the vortex does not always appear at the dot's center, which differs 
from the case in which similar dot on an infinite SC film. In closing, there
are two important contributions in semi-infinite HMSS; attraction of vortex to the edge 
through its image vortex and vortex-magnetization interaction. As a result of competion betweeen
these two factors, peculiar physical effects which do not come out in infinite HMSS, appear.
In this work, we studied the simplest cases to get idea about edge effects in HMSS. However, there
are still several interesting realizations that can be studied via the method that is developed here.    
We leave them for possible future works.

\end{multicols}


\begin{thebibliography}{10}

\bibitem{pok1} I.F. Lyuksyutov and V.L. Pokrovsky, Phys. Rev. Lett. 
{\bf
81}, 2344 (1998).

\bibitem{pl1} I.F. Lyuksyutov and V.L. Pokrovsky, in {\it 
Superconducting
Superlattices II: Native and
Artificial}, edited by Ivan Bozovic and Davor Pavuna, SPIE Proceedings
Vol. 3480 (SPIE-International Society for Optical Engineering, 
Bellingham,
WA,  1998), p. 230.

\bibitem{se1} S. Erdin, Physica C
{\bf 391},140 (2003).

\bibitem{pok2} I.F. Lyuksyutov and V.L. Pokrovsky, cond-mat/9903312.

\bibitem{pok3} I.F. Lyuksyutov and D.G. Naugle, Modern Phys. Lett. B {\bf
13}, 491 (1999).

\bibitem{e1} J.I. Martin, M. Velez, J. Nogues and I.K. Schuller, Phys.
Rev. Lett. {\bf 79}, 1929 (1997).

\bibitem{e2} D.J. Morgan and J.B. Ketterson Phys. Rev. Lett. {\bf 80},
3614 (1998).

\bibitem{e4} Y. Otani, B. Pannetier, J.P. Nozieres and D. Givord, J. Magn.
Mag. Mat. {\bf 126}, 622 (1993).

\bibitem{ee} M.J. Van Bael, K. Temst, V.V. Moshchalkov and Y.
Bruynseraede, Phys. Rev. B {\bf 59}, 14674 (1999).

\bibitem{vanbael} M.J. Bael, L. Van Look, K. Temst et al. Physica C
{\bf 332}, 12 (2000).


\bibitem{degennes} P.G. de Gennes, Sol. St. Comm. {\bf 3}, 127 (1965).

\bibitem{kramer} L. Kramer, Phys. Rev. {\bf 170}, 475 (1968). 

\bibitem{brandt} D.Y. Vodolazov, I.L. Maksimov and  E.H. Brandt, Physica C
{\bf 384}, 211 (2003).

\bibitem{domains} S.Erdin, I.F. Lyuksyutov, V.L. Pokrovsky and V.M.
Vinokur,
Phys. Rev. Lett. {\bf 88}, 017001
(2002).


\bibitem{th4} S. Erdin, A.M. Kayali, I.F. Lyuksyutov, and V.L.
Pokrovsky, Phys. Rev. B {\bf 66}, 014414 (2002).

\bibitem{abrikosov} A.A. Abrikosov, {\it Introduction to the Theory of
Metals}
(North Holland, Amsterdam, 1986).



\bibitem{kogan} V.G. Kogan, Phys. Rev. B {\bf 49}, 15874 (1994).

\bibitem{bean} C.P. Bean and J.D. Livingstone, Phys.Rev.Lett. {\bf 12}, 14 
(1964).
\end{thebibliography}
\end{document}